\documentclass[useAMS]{mn2e}

\usepackage{graphicx}

\title[Photoluminescence in amorphous MgSiO$_{3}$ silicate]{Photoluminescence in amorphous MgSiO$_{3}$ silicate}
\author[S. P. Thompson et al.]{S. P. Thompson$^{1}$\thanks{E-mail:
stephen.thompson@diamond.ac.uk}, J. E. Parker$^{1}$, S. J. Day$^{1,2}$, L. D. Connor$^{1}$,
A. Evans$^{2}$\\
$^{1}$Diamond Light Source, Harwell Science and Innovation Campus, Chilton, Didcot, Oxon OX11 0QX, UK\\
$^{2}$Astrophysics Group, Keele University, Keele, Staffordshire ST5 5BG, UK}

\begin{document}

\date{Accepted 1988 December 15. Received 1988 December 14; in original form 1988 October 11}

\pagerange{\pageref{firstpage}--\pageref{lastpage}} \pubyear{2002}

\maketitle

\label{firstpage}

\begin{abstract}
Samples of Amorphous MgSiO$_{3}$ annealed at temperature steps leading up to their crystallisation temperature show a rise in 
photoluminescence activity, peaking at $\sim$450 $\degr$C. The photoluminescence band has a main peak at 595 nm and a weaker 
peak at 624 nm. We present laboratory data to show that the maximum in photoluminescence activity is related to substantial 
structural reordering that occurs within a relatively narrow temperature range. We attribute the origin of the photoluminescence 
to non-bridging oxygen hole centre defects, which form around ordered nano-sized domain structures as a result of the breakup 
of tetrahedral connectivity in the disordered inter-domain network, aided by the loss of bonded OH. 
These defects are removed as crystallisation progresses, resulting in the decrease and eventual loss of photoluminescence. 
Thermally processed hydrogenated amorphous silicate grains could therefore represent a potential carrier of extended 
red emission.  
\end{abstract}


\section{Introduction}

Extended red emission (ERE) is a broad ($\Delta\lambda\sim$ 60-120 nm), featureless emission band with a peak wavelength between 600 
and 850 nm first detected in the spectrum of HD 44179 (the Red Rectangle; Schmidt et al. 1980). ERE is commonly seen in environments 
where both dust and UV photons are present (Witt \& Vijh 2004) and has been detected in reflection nebulae, dark nebulae, cirrus 
clouds, planetary nebulae, H\,{\sc ii} regions, novae, the diffuse interstellar medium (ISM), and the halos and high galactic latitude 
interstellar clouds of galaxies (e.g. Witt \& Schild 1988; Witt \& Boroson 1990; Scott et al. 1994; Szomoru \& Guhathakurta 1998; Smith \& 
Witt 2002; Rhee et al. 2007; Bern\'{e} et al. 2008; Witt et al. 2008). Darbon et al. (1999) showed that the peak wavelength 
and width of the ERE band are correlated, while Smith \& Witt (2002) showed that the peak wavelength also correlates with the density of the 
UV radiation field in a given ERE source. Examples of ERE illustrating the observational variation between different environments are 
shown in Fig. 1.

Since many solids emit visible luminescence when exposed to UV light, it is assumed that ERE is a photoluminescence (PL)\footnote{A 
list of uncommon acronyms is given in the appendix.} process 
powered by far-UV photons. In the ISM, $\sim$4$\%$ of the energy absorbed by dust at wavelengths $<$550 nm is emitted in the form of 
ERE, suggesting the carrier must be a major component of 
the interstellar grain population. The ERE carrier may have originally condensed in the circumstellar environment of proto-planetary 
and planetary nebulae and subsequently been ejected into the ISM (Witt et al. 1998). To survive the journey through the ISM and to 
spread throughout the halo of the Galaxy requires a robust carrier material. The intrinsic quantum yield of ERE may be as high 
as $\sim$50$\%$ with its carrier intercepting $\sim$20$\%$ of the photons absorbed by interstellar dust in the 90-550 nm range (Smith 
\& Witt 2002), which therefore limits the chemical composition of the ERE carrier to the few elements that are both abundant and 
highly depleted, i.e. C, Fe, Si and Mg. Since metals do not produce PL, the remaining possibilities are C and Si bearing solids.

\begin{figure*}
\includegraphics[height=5.5in, width=5.5in]{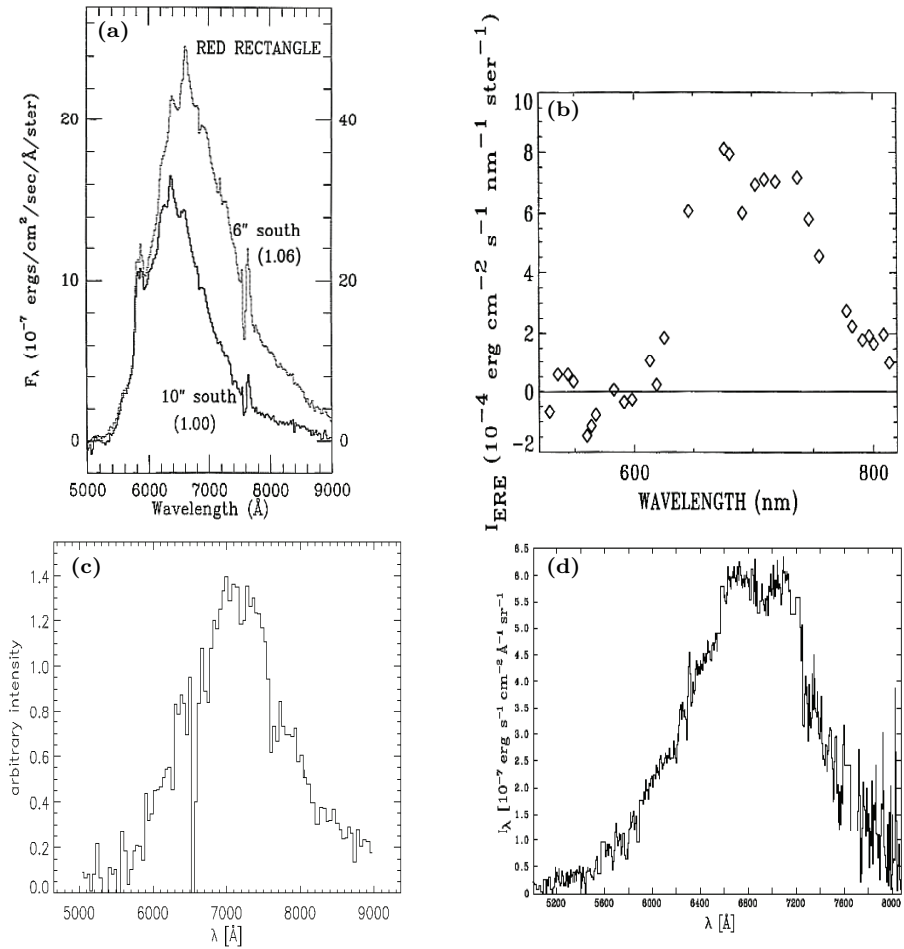}
\put(-335,365){\bf (a)}
\put(-165,335){\bf (b)}
\put(-345,162){\bf (c)}
\put(-165,162){\bf (d)}
\caption{Observed ERE for (a) the Red Rectange (from {\em Spectroscopy of extended red emission in reflection nebulae}, 
Witt, A. N., Boroson, T. A., Astrophysical Journal, 1990, 355, 182); (b) the planetary nebula NGC7027  (from {\em Extended red 
emission from dust in planetary nebulae}, Furton, D. G., Witt, A. N., 1992, Astrophysical Journal, 1992, 386, 587); (c) the NGC2327 reflection 
nebula (from {\em Extended Red Emission in the Diffuse Interstellar Medium}, Gordon, K. D., Witt, A. N., Friedmann, B. C. 
Astrophysical Journal, 1998, 498, 522) and (d) the north-western filament of the reflection nebula NGC 7023 (A. N. Witt 
personal communication, see also Witt et al. 2006). Figs. 1a-c reproduced by permission of the AAS.}
\end{figure*}

The physical nature of the ERE carrier is still contentious and various potential carriers have been proposed. These include 
hydrogenated amorphous carbon (Duley 1985; Duley \& Williams 1988; Witt \& Schild 1988; Furton \& Witt 1992; Duley et al. 1997; Seahra 
\& Duley 1999; Duley 2001; Godard \& Dartois 2010), quenched carbonaceous composites (Sakata et al. 1992), C$_{60}$ (Webster 1993), 
diamond (Duley 1988; Chang et al. 2006), embedded polycyclic aromatic hydrocarbon molecules (Wada et al. 2009), isolated 
dehydrogenated carbon clusters (Kurth et al. 2013) and silicon nanoparticles (Witt et al. 1998; Ledoux et al. 1998; 
Smith \& Witt 2002). In a 2003 review, Draine (2003) noted that, while a number of these proposed identifications have been 
ruled out, the arguments against them are not conclusive.  The fact that evidence for ERE was found in C-rich planetary nebulae 
(Furton \& Witt 1992) initially pointed to a carbonaceous carrier, but there appears to be no correlation between the 
3.3 $\mu$m UIR feature, generally attributed to PAH's, and the ERE in the Red Rectangle (Kerr et al. 1999); while amorphous carbon 
struggles to meet both efficiency and spectral constraints (Godard \& Dartois 2010). Later Infrared Space Observatory (ISO) 
observations of C-rich planetaries showed the presence of strong crystalline silicate features (Waters 
et al. 1998). However, ERE has not been observed in O-rich planetaries, appearing to make a straightforward association with 
O-rich materials difficult, particularly since the Red Rectangle, which is the most prominent ERE source with local dust production, 
is a C-rich environment.

Nevertheless silicates are a well known and abundant dust species whose presence in various environments is ubiquitous and Koike et 
al. (2002a,b; 2006) noted the similarity between the Red Rectangle ERE and thermoluminescence measured from gamma ray irradiated 
synthetic forsterite (Mg$_{2}$SiO$_{4}$). This gave an intense peak at 645--655 nm, while only very weak luminescence was observed in 
natural unirradiated olivine. However, other than these studies (and an apparently unpublished preprint by Koike et al. 2004 
suggesting a UV excited post-thermoluminescence PL effect in irradiated forsterite, attributed by them to nano-forsterite structures), there 
appears to have been little experimental or theoretical work examining the possibility of silicates as ERE carriers. 

Possibly related to the ERE phenomenon is the weak very broad-band structure (VBS) observed between 478 and 577 nm as a shallow depression 
on the interstellar extinction curve, first reported by Whiteoak (1966; see also Hayes et al. 1973; Shild 1977; van Breda \& Whittet 
1981; Krelowski et al. 1986). The VBS carrier has been proposed variously as an absorption effect (Hayes et al. 1973; Manning 1975;
Huffmann 1977), extinction by a population of very small amorphous carbonaceous grains (Jenniskens 1994), or as luminescence from hydrogenated 
amorphous carbon (Duley \& Whittet 1990).

In this paper we present PL measurements for samples of amorphous MgSiO$_{3}$ that have been annealed at temperatures leading up to 
crystallisation and show that their PL activity grows and peaks within a relatively narrow temperature range coinciding with the 
fragmentation of the silicate network that connects between nano-sized, but ordered domains. We suggest that silicate PL arises from 
non-bridging oxygen hole centres that form on the outer surfaces of the domain structures as a result of inter-domain 
network breakup and loss of bonded OH.

\section[]{Experimental}

\subsection{Sample manufacture}

Amorphous MgSiO$_{3}$ was produced by drying a sol-gel obtained by rapidly mixing 0.1 molar solutions of MgCl$_{2}$ and NaSiO$_{3}$: 
\[
\mbox{MgCl}_{2} + \mbox{NaSiO}_{3} + \mbox{H}_{2}\mbox{O} \rightarrow \mbox{MgSiO}_{3} + 2\mbox{NaCl} + \mbox{H}_{2}\mbox{O},
\]
the procedure for which has been described in detail elsewhere (Thompson et al. 2007, 2012). However in summary, following gelation 
the gels were washed and centrifuged several times in deionised water to remove the dissolved NaCl by-product. They were then dried in 
air at 75 $\degr$C for 24 hours. This results in large irregular glassy solids which were ground down to form a fine-grained powder. 
No attempt was made to control or select particle sizes, which after grinding, ranged from a few microns to several tens of microns in diameter, with 
a few approaching 50 to 100 $\mu$m.
Separate batches of this were then annealed at discrete temperatures between 100 $\degr$C and 700 $\degr$C using a 
Carbolite tube furnace. The annealing time at peak temperature for each sample was $\sim$17.5 $\pm$1 hours. All characterisation 
measurements were subsequently performed at ambient temperature.

\begin{figure}
\includegraphics[height=3.5in,width=3.5in]{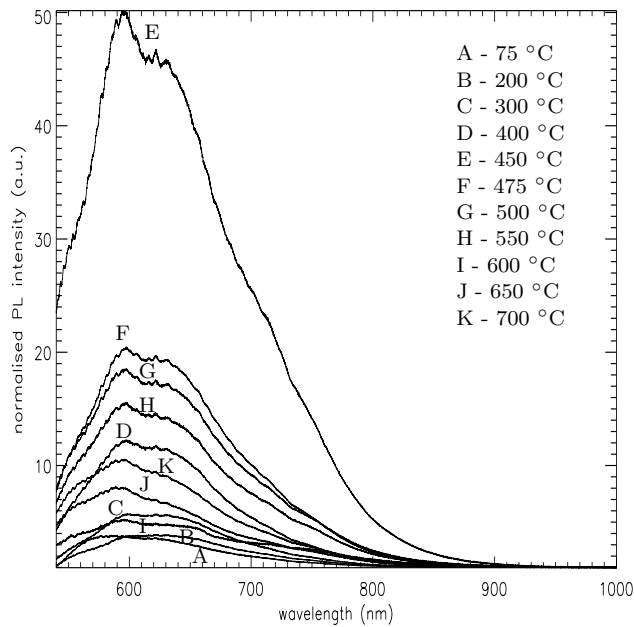}
{\footnotesize
\put(-170,30){A}
\put(-175,37){B}
\put(-202,48){C}
\put(-199,77){D}
\put(-188,228){E}
\put(-199,114){F}
\put(-190,100){G}
\put(-190,87){H}
\put(-190,41){I}
\put(-190,57){J}
\put(-183,64){K}
}
\put(-70,220){A - 75 $\degr$C}
\put(-70,210){B - 200 $\degr$C}
\put(-70,200){C - 300 $\degr$C}
\put(-70,190){D - 400 $\degr$C}
\put(-70,180){E - 450 $\degr$C}
\put(-70,170){F - 475 $\degr$C}
\put(-70,160){G - 500 $\degr$C}
\put(-70,150){H - 550 $\degr$C}
\put(-70,140){I - 600 $\degr$C}
\put(-70,130){J - 650 $\degr$C}
\put(-70,120){K - 700 $\degr$C}
 \caption{Normalised photoluminescence spectra for amorphous MgSiO$_{3}$ annealed at increasing temperatures between RT and 700 $\degr$C.}
\end{figure}

\begin{figure}
\includegraphics[height=3.5in,width=3.5in]{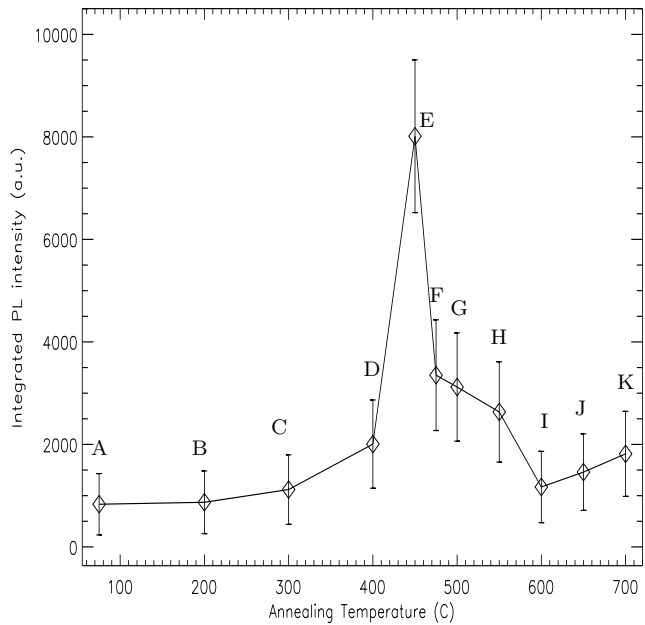}
\put(-218,70){A}
\put(-180,70){B}
\put(-150,78){C}
\put(-115,100){D}
\put(-94,194){E}
\put(-90,128){F}
\put(-82,123){G}
\put(-67,112){H}
\put(-48,80){I}
\put(-35,85){J}
\put(-19,95){K}
 \caption{Integrated intensity of the PL signal for amorphous MgSiO$_{3}$ as a function of annealing temperature. Labels A to K as per 
fig. 1.}
\end{figure}

\subsection{Data collection}
\subsubsection{Photoluminescence measurements}
Without further preparation, approximately equal quantities of powdered sample for each of the annealing temperatures were loaded on 
glass microscope slides such that each sample presented a roughly circular area $\sim$8 mm diameter and $>$0.5 mm thick. The surface 
was levelled off using the flat side of a stainless steel spatula. PL spectra were measured using a Horiba Jobin Yvon confocal LabRam 
800 spectrometer employing 532 nm laser excitation operated at 30 mW. A 600 line spectrometer grating was used in conjunction with a 
100 $\mu$m entrance slit and 300 $\mu$m confocal hole size; a 10$\times$ objective lens was used to both focus the incident laser spot (via 
imaging camera) and gather the PL radiation. In this configuration the laser spot on the sample was typically 300--400 $\mu$m in diameter, 
depending on the quantity of sample that lay in the focal plane. Indeed, as the overall sample surface was highly irregular due to the 
powder grains themselves being 
irregular in shape and orientation,  data were collected from at least ten randomly selected locations per sample and the PL spectra averaged. The
spectra per location were collected in 0.051 nm steps from 534 to 1000 nm in overlapping 50 nm segments, with each segment being the average 
of three 3 s collections. We did not observe any systematic size dependent effects, in that spots containing numerous smaller particles tended to 
give similar PL profiles to those with only a few larger ones. However for any given sample, the measured intensity often showed 
wide variations from spot to spot, due to various physical factors, including the uneven nature of the sample surface, particle orientation, size 
and possible variations in the internal homogeneity 
of the constituent particles in respect of the PL producing species. However, all measured spectra became asymptotically flat beyond $\sim$900 
nm and the intensity at 1000 nm (based on an average of the 20 data points between 999 and 1000 nm) was therefore used to normalise each 
individual spectrum prior to averaging.

\subsubsection{X-ray characterisation}
Materials with structural correlations that extend only over the nanometer length-scale give only broad diffuse X-ray scattering 
patterns rather than sharp Bragg diffraction peaks and represent the size domain where traditional X-ray crystallographic methods 
become less effective. However the total scattering (TS) method, also known as the pair distribution function (PDF) method, though 
similar to conventional powder diffraction, provides a means of extracting structural correlations at nano-scale distances from 
materials that do not possess long-range periodicities (see reviews by Proffen et al. 2003, 2006; Proffen \& Page 2004). Therefore, 
to supplement the structural analysis previously reported for these samples by Thompson et al. (2012, hereafter referred to as TPT, 
see summary in 
section 3.3), TS measurements were also made on three amorphous samples using beamline I12 at the Diamond Light Source 
synchrotron. These were the as-prepared sample dried at 75 $\degr$C and two that had been annealed at 300 $\degr$C and 600 $\degr$C 
respectively. These correspond to annealing temperatures at either side of the range where the strongest PL was observed and provide 
a ``before and after''  structural reference. Also for comparison, TS data were collected on a sample that had been extensively 
crystallised by annealing at 900 $\degr$C.

The main measurement criterion for TS compared to conventional powder diffraction is the need to measure to high magnitude values of 
the X-ray scattering vector, $Q$ ($=4\pi\lambda^{-1}\sin\theta$, where $\lambda$ is the X-ray wavelength and $\theta$ the incident 
angle). This captures information relating to short-range atom--atom distances and is essential for the resolution of 
atomic distances by Fourier inversion (see section 3.2). The need for measuring at high $Q$ and the weakness of the scattered signal 
in that region necessarily involves the use of high X-ray energies and relatively long exposure times, depending on X-ray source 
output characteristics. For X-rays of energy $E$ the maximum theoretically accessible $Q$ is $Q_{\rm max}=4\pi E/hc$ \AA$^{-1}$. In practice however 
$Q_{\rm max}$ is limited by the experimental geometry and detection arrangements, as is the minimum, $Q_{\rm min}$, value of $Q$ below 
which data cannot be collected (e.g. limited by the backstop arrangements necessary to prevent primary beam entering the detector).

For the TS measurements samples were loaded into 2 mm Kapton capillaries and scattering data collected using 87.4 keV monochromatic 
X-rays produced by a 4.2 T superconducting multi-pole wiggler insertion device in the Diamond storage ring. Data collection involved 
720 summed exposures per sample of 4 s each, acquired using a large pixellated 2D area detector (Thales Pixium 4343 with CsI scintillator 
on amorphous Si substrate: 148 $\mu$m$^{2}$ pixel size, 2880 $\times$ 2881 pixels; Daniels and 
Drakopoulos 2009). The detector was offset from the beam centre such that, at a distance of 500 mm from the sample, scattering data 
could be recorded out to $Q_{\rm max} = 30.8$ \AA$^{-1}$. To provide background correction, equivalent empty capillary and air scatter 
measurements were also made. The scattered intensity in the form of portions of the Debye--Scherrer rings captured on the area 
detector for each sample was integrated to give a two dimensional data set of intensity as a function of scattering angle (converted 
to $Q$), which was then corrected for background scatter and transformed to radial distribution functions, $G(r)$ (see next section), 
using the PDFgetX2 software\footnote{http://www.totalscattering.lanl.gov}. The X-ray energy (and precise sample--detector distance) was 
calibrated using the measured pattern from a standard reference sample of CeO$_{2}$ of known lattice parameter.

\section{Results and analysis}

\subsection{Photoluminescence measurements}

Fig. 2 shows the averaged PL spectra for each annealing temperature. The broad PL band shows a main peak at $\sim$595 nm (2.08 eV), 
with a weaker secondary peak at $\sim$624 nm (1.98 eV). The overall strength of the band exhibits a strong variation with annealing 
temperatures and Fig. 3 plots the integrated PL signal for each temperature step, showing a clear peak in the 
PL activity around $\sim$450 $\degr$C. The width of the error bars reflects the variation in measured intensity, due to both physical 
measurement factors (i.e. sample volume) and any intrinsic inhomogeneity in the distribution of PL centres within the silicate, as already 
discussed. However a clear rise in PL activity is observed above 200 $\degr$C, with a very narrow peak between 400 and 475 $\degr$C. Above this
the PL activity decreases back towards the base line level at 600 $\degr$C.

\subsection{Total scattering measurements}

The measured TS data are shown in Fig. 4. The as-prepared and the two samples annealed at 300 and 600 $\degr$C show TS patterns 
characteristic of scattering from non-crystalline structures (i.e. no sharp features), while the sample annealed at 900 $\degr$C shows 
well defined crystalline structure, evidenced by the appearance of sharp Bragg features. 

The real-space distribution, $G(r)$, of pairs of atoms within a material is described by
\begin{equation} 
G(r) = \frac{1}{r} \Sigma_{n}\Sigma_{m} \frac{f(0)_{n}f(0)_{m}}{\langle f(0)\rangle^{2}}\delta (r-r_{nm})-4\pi\rho_{0},
\end{equation}
where $\rho_{0}$ is the average atomic number density, $\delta$ is the Dirac delta function, $r_{nm}$ is the separation of 
the $n$'th and $m$'th atoms and the sums are over all atoms in the sample. The $f(0)$'s are the atomic form factors evaluated at 
zero magnitude of the X-ray scattering vector $Q$ and, to a good approximation, are given by 
the number of electrons, Z, 
on the atom. The angle brackets in the denominator indicate the average value. Thus, for a known atomic structure, $G(r)$ could be 
directly calculated and would consist of a series of delta functions, each at some distance $r$ corresponding to the separation 
distance between two atoms. The total pair distribution function (PDF) is built up by summing over all atom-pairs in the solid. 

However, $G(r)$ can be accessed experimentally through the measured TS pattern via Fourier transform,

\begin{equation}
G(r) = \frac{2}{\pi} \int_{Q_{\rm min}}^{Q_{\rm max}} Q[S(Q)-1]\sin (Qr)dQ.
\end{equation}
The $Q_{\rm min}$ and $Q_{\rm max}$ are defined by the experimental arrangement as described in section 2.2.2; and as wide a range in $Q$ as possible is 
required, particularly in respect of $Q_{\rm max}$, in order to provide adequate real-space resolution and to decrease termination ripples 
in $G(r)$ at short distances, caused by the finite $Q$ range.   

In Eq. (2), $S(Q)$ is the normalized powder diffraction pattern from the sample:

\begin{equation}
S(Q)=\frac{I^{coh}(Q)-\Sigma c_{i}|f_{i}(Q)|^{2}}{|\Sigma c_{i}f_{i}(Q)|^{2}}+1.
\end{equation}
$I^{coh}(Q)$ is the coherent scattering intensity, which is the measured powder pattern once it has been
corrected for experimental effects (background scattering, detector dead-time and efficiency), sample dependent effects (Compton and 
multiple scattering) and normalized for incident flux. The sums are over all the atomic species, $i$, present in the material with 
concentration $c_{i}$. Via this equation, $S(Q)$ and 
therefore $G(r)$ can be obtained from a suitably conditioned powder diffraction measurement. The definition of $S(Q)$ given above has 
the square of the atomic form factor in its denominator. This  becomes small for X-rays at high $Q$ and by dividing the measured 
intensity by this value there is a relative enhancement in the high $Q$ scattering, which contains the valuable short-range 
information ignored in conventional powder diffraction analyses.

\begin{figure}
\includegraphics[height=3.5in,width=3.5in]{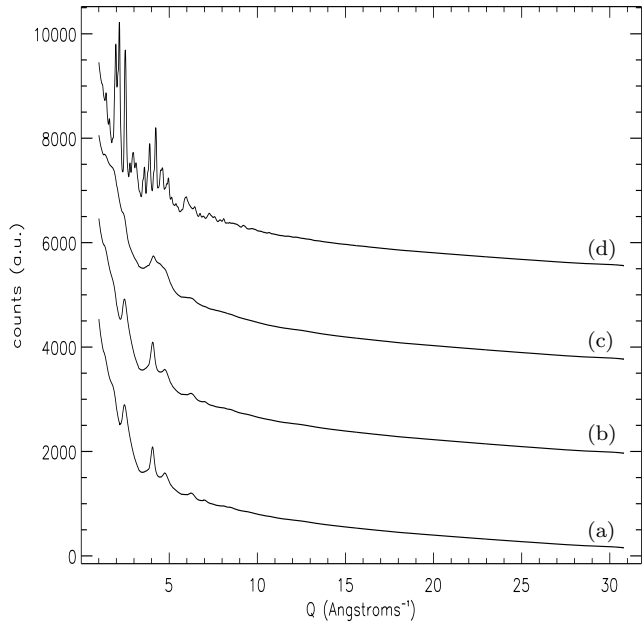}
\put(-30,145){(d)}
\put(-30,110){(c)}
\put(-30,75){(b)}
\put(-30,38){(a)}
 \caption{Measured X-ray total scattering diffraction patterns for amorphous MgSiO$_{3}$. (a) as manufactured (i.e. dried at 
75 $\degr$C); (b) annealed at 
300 $\degr$C; (c) 600 $\degr$C and (d) 900 $\degr$C.}
\end{figure}

In Fig. 5, the $G(r)$ radial pair distribution functions obtained by Fourier inversion are 
shown for the four samples. Each peak in $G(r)$ corresponds to a discrete atom pair distance and for the sample crystallised at 
900 $\degr$C atom to atom  correlations clearly extend beyond the 50 \AA\ shown in the plot. The three amorphous samples however 
show heavily damped $G(r)$ signals. The distance at which the oscillations in $G(r)$ diminish to zero represents the size of the 
coherent scattering domain (CSD) and is a measure of the distance over which the constituent atoms exhibit structural correlation. 
Both the as-manufactured and the 300 $\degr$C samples show a CSD size of $\sim$30 \AA, while in the 600 $\degr$C sample the CSD 
has reduced to $\sim$15 \AA, showing that there has been a reduction in the silicate's longer-range structural ordering as a direct 
result of annealing, that is, the silicate has become more amorphous.
 
There are two competing structural models for amorphous silicates: (i) the Zachariasen-Warren model (Zachariasen 1932, Warren 1933, 
1934, Warren et al. 1936) in which Si-O tetrahedra are linked together in a statistically disordered way to form a continuous random 
network with no long-range periodicity; and (ii) the microcrystal 
model proposed by  Lebiediev (1921), Randall et al. (1930a,b) and Valenkow \& Porai-Koshitz (1936). Here a continuous statistically 
random network provides links between the surfaces of ultrafine microcrystalline regions ($\sim$15-20 \AA) to form a domain structured 
material (Verweij \& Konijnendijk 1976). In Fig. 5, for distances less than $\sim$10 \AA, the $G(r)$ for all three samples shows 
very close similarities in the silicate structure in terms of the number, shape and positions of the component peaks, showing the 
amorphous silicate is described well by a nano-scale domain structured model.

\subsection{Summary of results from previous analyses}

The evolution of short- and medium-range structure leading up to crystallisation in these samples was investigated previously by TPT 
using FTIR spectroscopy at 10 $\mu$m, Raman spectroscopy and X-ray scattering at low $Q$. The Raman and X-ray data revealed that as 
the annealing temperature 
increases, there is a build up of strain within the silicate network, which is released  at $\sim$450 $\degr$C, causing a relaxation 
of both the Si--O--Si bond angle and Si-O bond length. Decomposing the 10 $\mu$m band at each temperature step allowed these changes 
to be related to changes in  the intertetrahedral connectivity. As implied by the stoichiometry of the MgSiO$_{3}$ composition, 
SiO$_{3}$ was found to be the initially dominant species and increased in proportion with rising annealing temperature as other 
tetrahedral species became incorporated into the SiO$_{3}$ chain structure. However in the region of $\sim$450 $\degr$C the 
proportion of SiO$_{3}$ rapidly decreased as species with greater numbers of non-bridging oxygen (NBO) atoms form. This 
coincides with a relaxation of the strain due to the breakup, or fragmentation, of the larger chain (and sheet) structures built from interconnected 
tetrahedra.

The X-ray analysis performed by TPT focussed on the scattering characteristics at low $Q$ (i.e. low scattering angle), however in 
Fig. 6 the conventional powder diffraction data collected by TPT is plotted for a wider angular range (equivalent to 0.2 to 
5 \AA$^{-1}$ in $Q$), showing qualitatively the evolution of the amorphous phase and formation of crystalline structure at 
$\sim$650 $\degr$C (pattern J in the figure). 
These data were recorded on the I11 beamline (Thompson et al. 2009) at the Diamond Light Source using flat-plate reflection geometry 
and low background, high resolution Si analyser crystals. The scattering features clearly visible in the 
5$\degr$ to 30$\degr$ 2$\theta$ range appear in the 2 to 7 \AA$^{-1}$ range in the TS patterns of Fig. 4, albeit with lower 
resolution due to differences in detection arrangements.  
Comparing Fig. 6 with Fig. 3 shows the peak in PL activity is clearly associated with the amorphous phase. The conclusion from 
the TS data that the 600 $\degr$C sample has become less ordered is also evident in the conventional powder pattern I, which despite 
the presence of a few exceedingly weak trace-level Bragg features which are below the detection level of the TS area detector, shows 
a loss of strength for the two prominent amorphous features at 14$\degr$ and 23$\degr$ 2$\theta$.

\begin{figure}
\includegraphics[height=3.5in,width=3.5in]{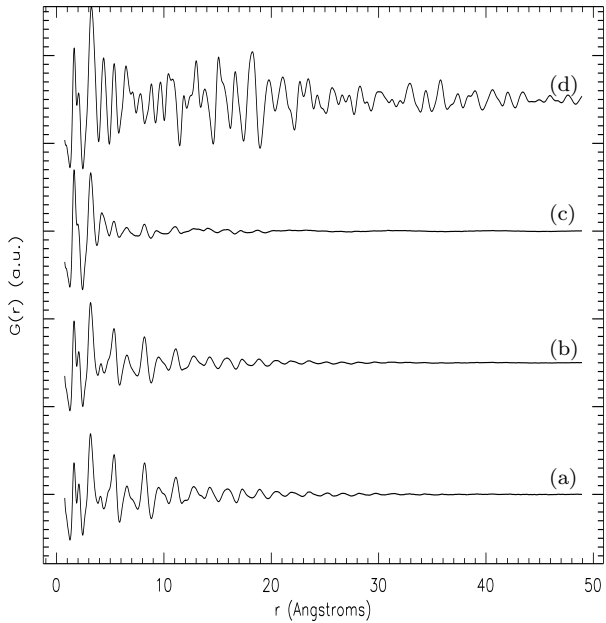}
\put(-30,207){(d)}
\put(-30,158){(c)}
\put(-30,107){(b)}
\put(-30,58){(a)}
 \caption{Radial distribution functions, $G(r)$, obtained from total scattering measurements for annealed amorphous MgSiO$_{3}$. 
Labels as per Fig. 4.}
\end{figure}

\begin{figure}
\includegraphics[height=3.5in,width=3.5in]{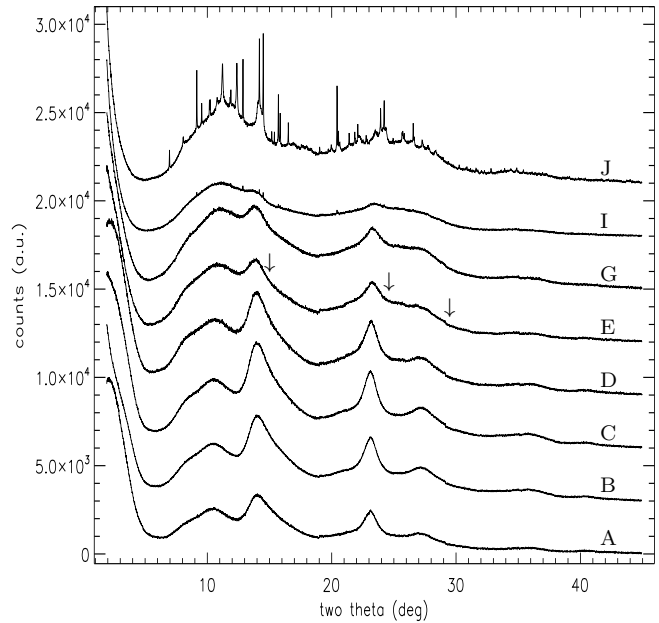}
\put(-30,35){A}
\put(-30,55){B}
\put(-30,75){C}
\put(-30,95){D}
\put(-30,115){E}
\put(-30,135){G}
\put(-30,155){I}
\put(-30,175){J}
\put(-157,140){$\downarrow$}
\put(-112,133){$\downarrow$}
\put(-89,123){$\downarrow$}
\caption{Conventional X-ray powder diffraction patterns for the MgSiO$_{3}$ samples measured using 15 keV X-rays at selected 
annealing steps. Plots are offset in y-axis for clarity, labels as per Fig. 2. The arrows shown for pattern E indicate the position
of where the three strongest diffraction lines of pure-phase silicon should occur if nanocrystalline silicon is present in the sample (see
section 4.1)}
\end{figure}

\section{Discussion}

The annealing temperature at which PL activity becomes strongest closely coincides with the temperature at which specific modifications in 
the form of structural fragmentation and changing of the non-bridging oxygen distribution were previously observed to occur; and suggests a 
relationship between structure and PL. Below we discuss four possible mechanisms for the observed PL: formation of Si nanoparticles, impurity doping, 
network deformation and non-bridging oxygen hole centres, concluding that non-bridging oxygen hole centres are 
the likely source of the PL.

\subsection{Embedded Si nanoparticles}

Si nanoparticles (SNPs) (Witt et al. 1998; Ledoux et al. 1998, 2000, 2001, 2002) with  1.5-5.0 nm diameters are widely considered to be
a strong contender for the ERE carrier (Witt \& Vijh 2004). PL in SNPs is believed to arise from a combination of quantum 
confinement in nanocrystals and the passivation of dangling surface bonds by H, O, N, C, or Fe atoms to inhibit non-radiative 
electron-hole recombination (Kovalev et al. 1999), though the details are still debated (e.g. Hannah et al. 2012). Laboratory studies 
involving O and H passivation show SNPs with H-passivation and diameters of 
$<$2.5 nm produce luminescence at blue and near-UV wavelengths, 
while O-passivation produces only red luminescence (Wolkin et al. 1999; Zhou et al. 2003). Because neither blue PL nor the Si-H 
absorption feature at 4.6 $\mu$m are observed under interstellar conditions, O-passivation is considered more probable (Witt et al. 
1998). 

Once produced in a circumstellar outflow O-passivated SNPs (O-SNPs) could persist, even in the ISM with abundant H atoms, since the 
Si-O bond is energetically two to three times stronger than the Si-H bond. However, such particles would also contribute to the 
10 $\mu$m silicate band via Si-O vibrations and stochastically heated O-SNPs should also produce a 20 $\mu$m emission band in excess 
of observationally established limits, while requiring an unrealistically high proportion of the Si abundance to be in the form of 
SNPs (Li \& Draine 2002). In addition, experiments indicate SNPs could lose their PL capability in environments where cosmic ion 
bombardment is significant (Baratta et al. 2004). These problems may be overcome if O-SNPs are either attached to, or embedded within, 
larger grains (Li \& Draine 2002; Witt \& Vijh 2004). Experiments involving embedded SNPs produced by ion implantation into solids 
show SNP luminescence can be preserved  (e.g. Iwayama et al. 2002), but may be red-shifted (relative to PL from ``free-standing'' 
SNPs) by compressive strain imparted by the embedding matrix (K\r{u}sov\'{a} et al. 2012). X-ray absorption near edge structure 
(XANES) spectroscopy at the Si K-edge show the average environment surrounding the Si atoms in these silicates prior to annealing 
(i.e. weak PL activity) is dominated by medium-range tetrahedral Si-O symmetry (Thompson 2008) rather than cubic Si-Si environments. 
Fig. 7 compares the first 8 \AA\ of the $G(r)$ obtained for the sample annealed at 600 $\degr$C (i.e. weak post-annealing PL) with 
theoretical $G(r)$'s calculated using the DISCUS diffuse scattering simulation software (Neder \& Proffen 2008) for 2 nm spheres 
of crystalline silicon (Fd3m structure and 5.4307 \AA\ lattice parameter) and crystalline enstatite (Pbca structure, lattice 
parameters 18.216, 8.813, 5.179 \AA). Although the experimentally derived $G(r)$ contains less fine structure, reflecting its more 
disordered nature (crystallographic Debye-Waller temperature factors were used in the simulations), the general peak positions 
are quite well accounted for by the silicate structure. Peaks at 2.35 and 3.83 \AA\ corresponding to the first and second Si-Si 
distances are clearly absent from the experimental $G(r)$ (and are similarly absent in the other experimentally derived $G(r)$'s). 

However, there are reports in the literature regarding SNP PL in crystalline and amorphous SiO$_{2}$ induced by  
Si implantation (e.g. Linnros et al. 1999), so we might consider the possibility that thermally induced network fragmentation might 
produce SNPs, that are naturally O-passivated, via the formation and linking together of structural units with one or more dangling Si 
bond, somewhat akin to the Si/SiO$_{2}$ core mantle nano grains that Li and Drain (2002) suggested could exist as ERE carriers, either 
as $>$50 \AA\ clusters or as attachments to larger grains. The weak PL activity above $\sim$600 $\degr$C however would presumably require their 
``reabsorption'' back in to the silicate structure at higher annealing temperatures, which is plausible as the Si-Si bond dissociation 
energy is approximately half that of the Si-O bond.  Indicated in Fig. 6 are the locations where 
the first three strongest peaks of crystalline cubic Si (shown by arrows at 15.1$\degr$, 24.5$\degr$ and 29.4$\degr$ 2$\theta$) 
should appear in the conventional powder diffraction pattern for the sample annealed at 450 $\degr$C. For 2 nm Si particles the expected 
diffraction peak width will be at least 2$\degr$ in 2$\theta$, calculated via the well known Scherrer equation 
$B(2\theta)=0.94\lambda/D\cos(\theta)$, which relates crystallite size $D$ to the width $B$ of a reflection of scattering angle $\theta$.
For much larger particles, any such peaks should become correspondingly narrower, however 2 nm represents a significant proportion of the
CSD derived from the TS measurements and the formation of SNPS in any significant number should have an observable effect on the morphology of
the conventional powder pattern. No such features are apparent in Fig. 6, suggesting SNPs do not, or cannot, form by the thermal annealing of 
amorphous silicate.

\begin{figure}
\includegraphics[height=3.5in,width=3.5in]{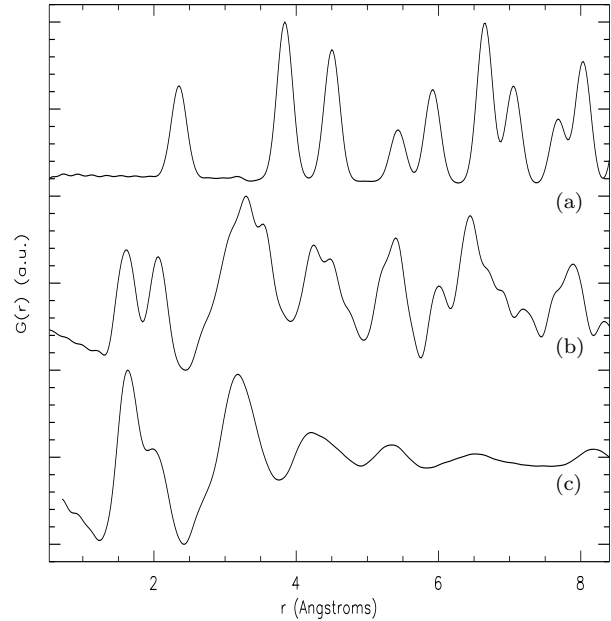}
\put(-30,162){(a)}
\put(-30,107){(b)}
\put(-30,55){(c)}
 \caption{Calculated radial distribution functions, $G(r)$, out to 8\AA\ for (a) bare 2 nm diameter Si particles and (b) 2 nm 
silicate particles with enstatite structure compared with (c) measured $G(r)$ for the amorphous silicate annealed at 600 $\degr$C.}
\end{figure}

\subsection{Impurity doping}

Due to its potential technological importance (e.g. lasers Petri\v{c}evi\'{c}  et al. 1988), PL in doped crystalline forsterite 
(Mg$_{2}$SiO$_{4}$) and enstatite (MgSiO$_{3}$) have received much interest. Of relevance to the present discussion are Mn impurities 
since these produce luminescence lines at 630 and 670 nm in forsterite and enstatite respectively (references in  MacRae \& Wilson 
2008) and submicron Mn-rich
forsterite and enstatite crystals have been found in chondritic, porous IDPs, the matrices of chondrite meteorites (Kl\"{o}ck et al. 
1989; Vollmer et al. 2009) and in olivine in comet 81P/Wild 2 nucleus samples (Zolensky et al. 2006), though olivine with low Fe and 
elevated Mn is thought to have formed from
condensation in the protosolar nebula (Kl\"{o}ck et al. 1989). The use of high purity reagent grade starting materials 
(MgCl$_{2}$ $>$99$\%$ Fisher Scientific and NaSiO$_{3}$ Fluorochem) in the sol-gel procedure, although used as supplied, should rule 
out impurities as the source of PL. Unless annealing provides specific, but thermally temporary, structural sites for PL to be produced by 
impurities, any such species would be present in all samples over the temperatures studied, as should their PL. Furthermore, X-ray 
fluorescence spectra (Horiba XGT-7000 X-ray Fluorescence Microscope) collected  from different regions of a portion of the as-prepared 
sample showed no lines in the region of 6 keV characteristic of Mn K$_{\alpha}$ fluorescence. The observation by 
Koike et al. (2002a,b 2006) that luminescence can be induced in forsterite that has been structurally disrupted by gamma rays likely 
points to an intrinsic structure related mechanism.

\subsection{Network deformation}

Shinno et al. (2000) observed luminescence in crystalline forsterite subjected to shock pressures in the range of 10 to 82 GPa. 
Correlating shifts in the Mg-O translational and Si-O rotational Raman bands at 308 and 328 cm$^{1}$ with the growth of luminescence 
intensity as a function of pressure, they attributed the luminescent mechanism to lattice deformation, which allows transitions to 
occur between deformed electron levels and vibrational sublevels in the [SiO$_{4}$]$^{4-}$ molecular orbital (Shinno et al. 1999). 
In TPT shifts to higher frequency in the position of the 670 cm$^{-1}$ Raman band, attributable to a narrowing of the Si-O-Si 
intertetrahedral bond angle, $\phi_{T-T}$, and a stretching of the Si-O bond length, $r_{Si-O}$ (also deduced by TPT from X-ray 
scattering), indicate that the silicate structure becomes increasingly strained with annealing temperature. Fig. 8 reproduces the 
TPT Raman data. Clearly visible is a general trend towards higher frequency, but with anomalously large shifts in the region of 
450 $\degr$C. The sudden decrease in frequency between 475 and 500 $\degr$C represents the release of built up strain by structural 
re-ordering and changes in the relative abundances of SiO$_{2}$, Si$_{2}$O$_{5}$, SiO$_{3}$, Si$_{2}$O$_{7}$ and SiO$_{4}$ 
species within the silicate (see TPT for detailed discussion). The peak in PL activity also coincides with the peak in strain at
 $\sim$450 $\degr$C, suggesting a possible link. However, the luminescence observed by Shinno et al. peaked at $\sim$400 nm, with a 
long tail decay showing little activity by 600 nm and was the result of high pressure deformation, which seems unlikely to be 
replicated by the relatively low temperatures used here. Furthermore forsterite, although known to be the first crystalline phase to 
form in amorphous MgSiO$_{3}$ (Rietmeijer et al. 1986; Thompson \& Tang 2001; Roskosz et al. 2009), contains isolated SiO$_{4}$ 
tetrahedra while the 670 cm$^{-1}$ Raman band relates to changes in the intertetrahedral bond angle $\phi_{T-T}$ and therefore strain 
within linked tetrahedral structures. However, the close occurrence of the two peaks in PL activity and strain in the region of 
450 $\degr$C means that we cannot necessarily rule out a deformation-based contribution to the PL (see  section 4.5). 

\subsection{Non-bridging oxygen hole centres (NBO-HC)}

Due to its industrial and commercial importance, most studies of luminescence defect centres in undoped silicate focus on SiO$_{2}$, 
where four intrinsic types have been identified: (i) the {\it E$\,^{\prime}$} centre due  to either a paramagnetic positively charged 
oxygen vacancy $\equiv$Si$\cdot$Si$\equiv$ (where $\equiv$Si represents the bonding to three other oxygen atoms in the Si-O 
tetrahedron), or to a neutral dangling silicon bond  $\equiv$Si$\cdot$; (ii) a dangling oxygen bond $\equiv$Si-O$\cdot$ which 
is the NBO-HC; (iii) a peroxy radical (POR) $\equiv$Si-O-O and (iv) oxygen deficiency centre (ODC) due either to a neutral oxygen 
vacancy $\equiv$Si-Si$\equiv$ or twofold-coordinated silicon atom $\equiv$Si-O-Si-O-Si$\equiv$. Of these only the POR and NBO-HC 
defects produce PL in the region of 2 eV (Skuja 1998). 

POR defects can be formed by the combination of $\equiv$Si$\cdot$ {\it E$\,^{\prime}$} centres with interstitial O$_{2}$ molecules 
(Edwards \& Beall-Fowler 1982). Re-examination of the TPT Raman data collected at each annealing step shows no Raman feature in the 
region of 1549 cm$^{-1}$, that would be characteristic of interstitial O$_{2}$ (Skuja et al. 1998). The rise in PL with annealing is 
therefore unlikely to be due to POR defects.

Deconvolution of the mid-IR 10 $\mu$m band by TPT showed the strain release peak at $\sim$450 $\degr$C to coincide with a 
significant change in the relative proportions of the various tetrahedral species within the silicate, while changes in the silicate's IR 
hydration band near 3 $\mu$m suggested this was likely driven or facilitated 
by dehydrogenation of the sample. Consequently, the PL may be related to the formation of defects in the silicate structure resulting 
from this process. 

In naturally occurring terrestrial silicate minerals  single, or multi-step, dehydroxylation occurs in the range 
500-700 $\degr$C. However, The specific dehydroxylation temperature is determined primarily by the mineral's specific crystal structure and 
it's octahedral and extra-framework cation compositions, with various intermediary crystal phases forming as dehydroxylation 
progresses (e.g. Che et al. 2011 and references therein). Furthermore, the  dehydroxylation temperature is 
lowered if the mineral structure becomes disordered (Tyburczy \&\ Ahrens 1988) such that the structural changes seen here and by TPT 
are likely to be related to the loss of bonded OH. In amorphous SiO$_{2}$ PL bands are observed at 629 nm (1.97 eV) and 590 nm 
(2.1 eV) and are  due, respectively, to the following reactions (Munekuni et al. 1990):

\begin{equation}
\equiv \mbox{Si-O-O-Si}\equiv + \;e^{-} \rightarrow\; \equiv\mbox{Si-O}\!\uparrow\, ^{-}\mbox{O-Si}\equiv
\end{equation}
and
\begin{equation}
\equiv \mbox{Si-O-H H-O-Si}\equiv \;\rightarrow\; \equiv\mbox{Si-O}\!\uparrow\cdots\mbox{H-O-Si}\equiv +\;\mbox{H}^{0},
\end{equation}

In reaction (4) the $\equiv$Si-O$^{-}$ anion, once excited, acts as an energy donor, transferring its excess energy to the NBO-HC energy 
acceptor defect that has been formed together with the $\equiv$Si-O$^{-}$. The NBO-HC is them pumped to the excited state, returning to 
the ground state by luminescence. Reaction (5) on the other hand requires the existence, or formation, of  closely located $\equiv$Si-O-H 
pairs from which a hydrogen atom is released to leave a hydrogen bonded NBO-HC. The two peaks in the PL spectrum at 595 and 624 nm (2.08 
and 1.98 eV respectively) of our samples are very close to the SiO$_{2}$ NBO-HC PL peak energies and therefore suggest a similar NBO-HC 
origin. To support this, Fig. 9 compares PL spectra measured for samples of commercially sourced silica 
(amorphous SiO$_{2}$), mineral enstatite and two mineral olivines. All three exhibit clear PL features, albeit with weak activity, close to 590 and 630 nm and similar to those
observed in the spectra of the annealed amorphous silicate. 

Fig. 10 plots the percentage of NBO atoms in the silicate for each annealing temperature. These are derived from the relative 
proportions of the SiO$_{3}$, Si$_{2}$O$_{7}$, Si$_{2}$O$_{5}$ 
and SiO$_{4}$ obtained by TPT by decomposition of the 10$\mu$m band. There is a $\sim$10 $\%$ increase in the NBO content between 
200 $\degr$C and 475 $\degr$C which, taken in combination with the reduction in the CSD size, implies a breaking of intertetrahedral 
connections. Since the core structure of the CSD remains unchanged this can be viewed either as a ``surface" 
fragmentation of the structured domains, or as a growth and penetration of the random interdomain network into the ordered domains. We 
suggest therefore that during this process NBO-HC defects, associated with newly formed NBO sites around the domain 
surface, give rise to the observed PL. 

\begin{figure}
\includegraphics[height=3.5in,width=3.5in]{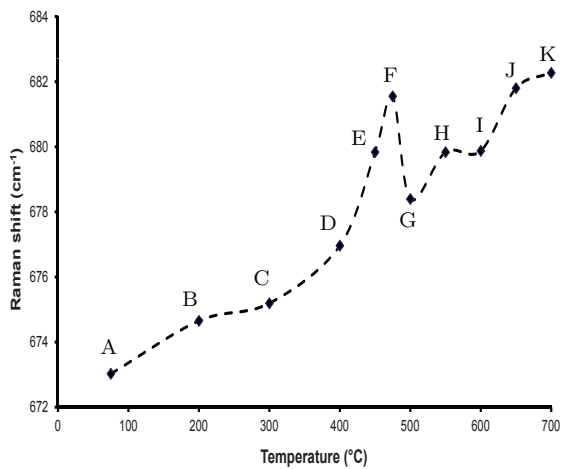}
\put(-196,82){A}
\put(-165,100){B}
\put(-138,108){C}
\put(-113,129){D}
\put(-101,161){E}
\put(-89,185){F}
\put(-83,130){G}
\put(-70,163){H}
\put(-54,166){I}
\put(-43,187){J}
\put(-30,193){K}
 \caption{Shift in peak position of MgSiO$_{3}$ 670 cm$^{-1}$ medium-range order Raman feature as a function of annealing temperature. 
Shifts towards higher frequencies are characteristic of a narrowing of the Si-O-Si bond angle and a shortening of the Si-O bond 
distance. The sharp decrease between 475 and 500 C is attributed to strain release. Labels as per Fig. 2.}
\end{figure}

\begin{figure}
\includegraphics[height=3.5in,width=3.5in]{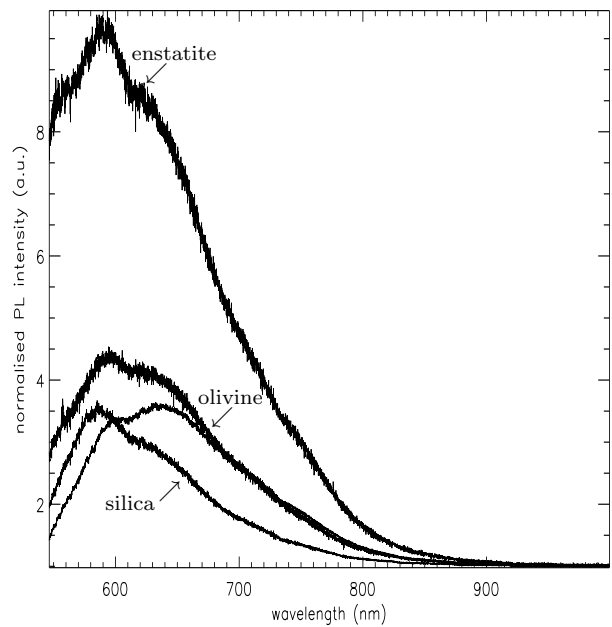}
\put(-200,50){silica}
\put(-180,55){$\nearrow$}
\put(-165,90){olivine}
\put(-160,83){$\swarrow$}
\put(-190,220){enstatite}
\put(-185,213){$\swarrow$}
\caption{PL measured for Silica, mineral enstatite and two olivine mineral samples.}
\end{figure}

\begin{figure}
\includegraphics[height=3.5in,width=3.5in]{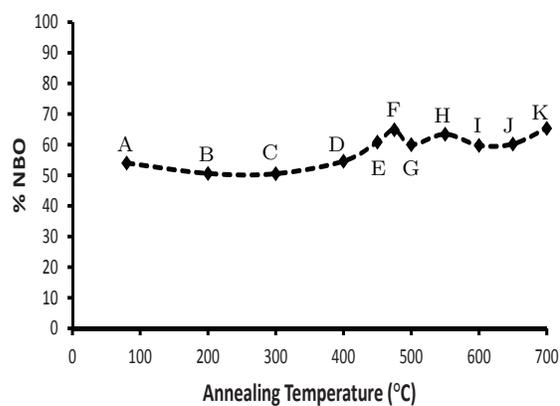}
\put(-189,145){A}
\put(-158,141){B}
\put(-134,142){C}
\put(-109,145){D}
\put(-93,136){E}
\put(-87,158){F}
\put(-81,136){G}
\put(-69,156){H}
\put(-54,152){I}
\put(-43,152){J}
\put(-32,157){K}
 \caption{Percentage of non-bridging oxygen atoms in the MgSiO$_{3}$ silicate at each annealing temperature. Labels as per Fig. 2.}
\end{figure}

\subsection{Astrophysical relevance}

Due to their formation in H-rich environments, bonded hydrogen defects such as SiH and SiOH have long been believed to be present in 
cosmic silicates (e.g. Moore et al. 1991; Whittet et al. 1997; Timmermann \&\ Larson 1993; Malfait et al. 1999; Thompson et al. 2003) 
though the close proximity of O-H features from water ice have hindered observational identifications. Hydroxylated amorphous silicates on 
the other hand have been indentified in interplanetary dust particles (IDPs) and glasses with embedded metal and sulphides (GEMS; 
Thomas et al. 1993; Bradley 1994; Bradley et al. 2005) some of which may be pre-solar (Messenger et al. 2003) and originate from the 
ISM (Bradley et al. 1999). The presence of OH groups in these recovered amorphous silicates is thought to result 
from hydrogen implantation by irradiation processes (Bradley 1994); while laboratory work by Djouadi et al. (2011) shows stable OH defects in ISM 
silicates should result from low energy proton irradiation in shock waves. 

In a crystalline silicate the total number of hydroxylated sites is small due to the limited surface area of the crystal edge regions where 
OH can form, thus limiting the number of potential NBO-HC sites and hence yield only the low levels of PL activity observed in mineral species. 
In amorphous silicates however the inherent disorder, possibly further supplemented by the disruptive effects of ion implantation 
(Demyk et al. 2001, 2004; Brucato et al. 2003; J\"{a}ger et al. 2003; Bringa et al. 2007; Davoisine et al. 2008)  should provide for greater
numbers of OH edge defects capable of producing NBO-HC's when annealed and, hence, an increase in PL activity. The level of PL produced by silicate 
grains in a potential ERE environment will thus depend on whether the grains have experienced (a) sufficient hydrogenation to produce bonded OH 
and (b) sufficient thermal processing at some point to convert these to NBO-HC's (and would not necessarily have to have occurred in situ within 
the ERE emitting environment). 
Observationally, silicate PL should thus correlate with weakened, or absent, silicate OH features in the mid-IR.

\begin{figure}
\includegraphics[height=3.0in,width=3.5in]{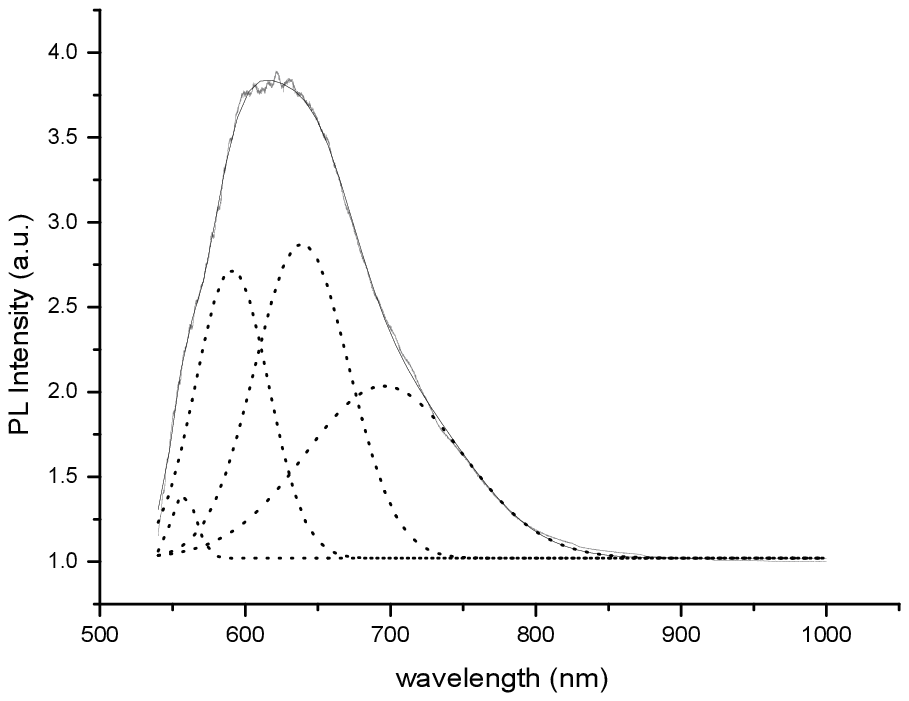}
\put(-200,190){(a)}
\qquad
\includegraphics[height=3.0in,width=3.5in]{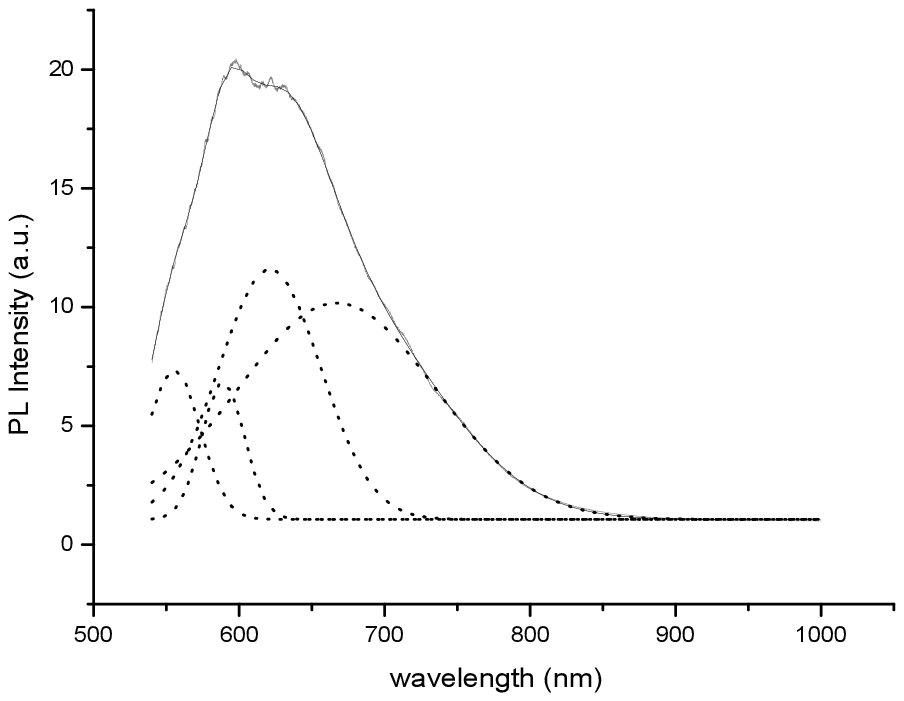}
\put(-200,190){(b)}
 \caption{Gaussian decomposition for the  PL spectra for amorphous  MgSiO$_{3}$ (a) annealed at 200 $\degr$C and (b) annealed at 475 $\degr$C. Component 
peaks for (a) are located at 557.64, 590.91, 639.00 and 695.42 nm (2.223, 2.098, 1.940 and 1.783 ev respectively) and for (b) are located at 
554.07, 587.54, 621.03 and 659.99 nm (2.238, 2.110, 1.996 and 1.879 eV respectively).}
\end{figure}

The change in the relative distribution of tetrahedral species observed by TPT and the domain structure reported here raises the 
possibility that NBO-HC defects could form in different regions of the silicate structure rich in one or other of the Si-O species. 
Indeed the large error bars in the integrated PL intensity, particularly for the 450 $\degr$C sample, arise from the increased variation in the 
measured PL signal obtained for different spots on the powder sample; and likely indicates phase inhomogeneity caused, or exacerbated, by annealing. 
These slightly different structural environments could offer a means of tuning both the PL band width and its peak wavelength in different 
environments. In 
Fig. 11 we show the results of a Gaussian decomposition of the PL bands measured for the samples annealed at 200 $\degr$C and 475 $\degr$C (label F in Figs. 2 
and 3). Four components were necessary to reproduce the measured signal and were located at 554.07, 587.54, 621.03 and 659.99 nm 
(2.24, 2.11, 1.99 and 1.88 eV respectively). A similar fit to the 200 $\degr$C band gave component positions of 559.52, 593.27, 632.70 
and 641.87 nm (2.22, 2.09, 1.96 and 1.93 eV respectively) which, with reference to Fig. 8 represent points of high and low strain. 
In both cases the energies are sufficiently close to the energies of reactions 4 and 5 to suggest that four distinct NBO-HC 
environments may be contributing to the PL. That the relative numbers of these can vary is also apparent in Fig 9, where the weak shoulder at 
$\sim$554 nm visible in the annealed samples in Fig. 1 is more pronounced in the mineral enstatite spectrum compared to either the olivine or silica spectra. 
On the other hand, the weaker $\sim$659 nm feature in both the annealed samples and the enstatite and silica spectra is more prominent in one of the
olivine spectra. As is the case with embedded SNPs these small variations in energy may be related to strain in the surrounding silicate network 
in which the two generic NBO-HCs are embedded, which could affect the local symmetry to possibly produce four distinct sites, or contribute directly 
to their formation (Hibino \& Hanafusa 1988). 

From the foregoing, the narrow temperature range over which the PL activity peaks appears to be clearly related to the thermally driven 
evolution of an initially hydrated amorphous structure, with initial changes in the number of NBOs allowing PL active NBO-HCs to form. As these structural 
changes continue towards crystallisation, the PL activity decreases as the NBO-HC defects become ``reabsorbed'' back into the silicate structure. 
Therefore, increased PL activity is a characteristic of amorphous grain processing prior to crystallisation. As such, ERE due to silicate PL should 
be anti-correlated with IR features due to crystalline grains. The range in temperature reported here over which PL is observed however could be 
widened if the PL-reducing effects of thermal annealing are offset by structural disruption and rehydrogenation due to ongoing ion implantation. 

Comparing Figs. 1 and 2 suggests that the silicate PL, peaking at $\sim$650 nm, better matches the ERE from the Red Rectangle protoplanetary 
nebula (Fig. 1a) rather than the ERE from the two reflection nebulae (Fig. 1c and 1d), where the peak wavelength is shifted longward to $\sim$700 nm, 
though the low resolution spectrum for the NGC7027 planetary nebula (Fig 1b) is suggestive of a  component close to 650 nm. One possibility for the peak 
position variability observed from source to source is that ERE does not originate from a single specific carrier, but rather a set of different carriers 
whose specific PL producing properties (e.g. relative abundance, size distribution etc.) is determined by local conditions (A. N. Witt, personal 
communication). In this respect, silicate PL at $\sim$650 nm may represent one component of a wider ERE producing grain population and may contribute 
to a lesser or greater extent in different sources.

Finally, silicate PL via NBO-HC's should not produce excess contributions to the mid-IR 10 and 20 $\mu$m bands as, unlike O-SNPs, an additional 
vibrating Si-O species is not invoked. It may be possible however that luminescence from silicates could be observable at other wavelengths. Koike 
et al. (2006) observed blue-shifted thermoluminescence in samples of forsterite and fused quartz that had been neutron irradiated. Although the 
luminescence mechanism was not identified, peak emission occurred between 340 -- 500 nm and 420 -- 520 nm, respectively, both of which are close, for example, 
to the interstellar VBS observed between 478 and 577 nm. We may therefore speculate that the formation of stable defects in silicate grains, akin to those outlined in
section 4.4, might also give rise to other observable emissions outside of the ERE region. 

\section{Conclusions}

We have observed the rise and decline of PL activity in thermally processed  amorphous MgSiO$_{3}$ and identify this with the formation 
and reabsorption of non-bridging oxygen hole centres which form when intertetrahedral connections are broken and reformed as a result 
of structural rearrangement and dehydroxylation. Decomposition of the PL band suggest $\sim$4 distinct defects are contributing to the 
observed PL signal, which we interpret as being due to hole centres associated with regions of differing tetrahedral connectivity, while 
differences in their PL energies may be related to distortions of the hole centre environment by strain within the silicate network. 
X-ray scattering shows ordered nanocrystalline domains to be present within the silicate, whose size is reduced with annealing, 
suggesting that the non-bridging oxygen hole centres most likely form around the shrinking surface of the nanocrystal domains as it 
becomes incorporated into the random network structure that links between the domains. As crystallisation proceeds, the hole centres 
become reabsorbed into the structure and PL activity decreases. The formation of non-bridging oxygen hole centres requires the 
presence of hydrogen atoms in the form of bonded OH molecules and can be achieved either by formation in H-rich environments and/or by 
implantation.  If astronomical red PL, i.e. ERE, originates (or at least in part) from silicates, then its presence will be indicative of hydrated grains 
having, at some point in their history, undergone low-level thermal processing; while its absence may indicate anhydrous grains and/or grains that have undergone
extensively thermal processing. Differences in the PL emission characteristics may derive naturally from differences in structural 
state, compositional variation or differing H implantation conditions.

\section*{Acknowledgments}

This work was supported by Diamond Light Source through beamtime award 7124. SJD acknowledges financial support from Diamond Light 
Source and Keele University. The authors would also like to extend their thanks to Professor A.N. Witt for providing the ERE data 
for NGC7023 and additional discussions regarding ERE; and to Professor L. d'Hendecourt for his helpful comments and suggestions during review.

\section*{Appendix}

\subsection*{Quantities and definitions}                      
\begin{tabular}{l l} \hline
CSD     & coherent size domain \\
ERE     & extended red emmission \\
{\it E$\,^{\prime}$} & colour centre due to missing oxygen atom \\
GEMS    & glasses with embedded metal and sulphides \\
$G(r)$  & atomic pair distribution function (PDF) \\
NBO     & non-brdiging oxygen (atom) \\
NBO-HC  & non-bridging oxgyen hole centre \\
ODC     & oxygen deficiency centre \\
O-SNP   & oxygen passivated silcon nano particle \\
PDF     & (atomic) pair distribution function ($G(r)$)\\
$\phi_{T-T}$ & inter-tetrahedral bond angle \\
PL      & photoluminescence \\
POR     & peroxy radical \\
$Q$     & X-ray scattering vector \\
$r_{Si-O}$ & silicon to oxygen bond length \\
$S(Q)$  & normalised scattered intensity \\
SNP     & silicon nano particle \\
TPT     & Thompson, Parker \& Tang (2012) \\
TS      & total (X-ray) scattering \\
VBS     & very broad-band structure \\
XANES   & X-ray absorption near edge structure \\

\hline                                
\end{tabular}

\bsp

\label{lastpage}

\end{document}